\documentclass[review,1p,authoryear,10pt]{elsarticle}
\usepackage{graphicx}
\usepackage{lineno}
\usepackage{amsmath,amssymb}
\usepackage{natbib}
\usepackage{rotating}
\usepackage{epstopdf}
\DeclareGraphicsExtensions{.jpg,.pdf,.mps,.png}

\newcommand{\vect}[1]{{\bf #1}}

\newcommand{\eqi}{\begin{equation}}
\newcommand{\eqo}{\end{equation}}
\newcommand{\pict}[4][5cm]{
\begin{figure}
\centerline{\includegraphics[width=#1]{./#2}}
\caption{\footnotesize{#3}}
\label{#4}
\end{figure}}
\newcommand{\zf}{\left < U_{\phi} \right >}

\begin{document} 
\title{Experimental study of libration-driven zonal flows in non-axisymmetric containers}
\author{J. Noir}
\address{ETH Z\"urich, Institut fur Geophysik, Sonneggstrasse 5 - CH8092 Z\"urich - Switzerland \\ 
Department of Earth and Space Sciences, University of California, Los Angeles, CA 90095-1567  USA}
\ead{jerome.noir@erdw.ethz.ch}
\author{D. C\'ebron, M. Le Bars, A. Sauret}
\address{Institut de Recherche sur les Ph\'enom\`enes Hors \'Equilibre, CNRS and Aix-Marseille University, 49 rue F. Joliot-Curie, BP 146 -
F13384, Marseille cedex 13 - France}
\author{J.M. Aurnou}
\address{Department of Earth and Space Sciences, University of California, Los Angeles, CA 90095-1567  USA}

\begin{frontmatter}
\begin{abstract}
Orbital dynamics that lead to longitudinal libration of
celestial bodies also result in an elliptically deformed equatorial
core-mantle boundary. 
The non-axisymmetry of the boundary leads to a topographic coupling between the assumed rigid
mantle and the underlying low viscosity fluid.
The present experimental study investigates the
effect of non axisymmetric boundaries on the zonal flow driven by longitudinal libration. For large enough equatorial ellipticity, we report intermittent
space-filling turbulence in particular bands of resonant frequency correlated with larger amplitude zonal flow. The mechanism underlying the intermittent turbulence has yet to be unambiguously determined. Nevertheless, recent numerical simulations in triaxial and biaxial ellipsoids suggest that it may be associated with the growth and collapse of an elliptical instability \cite[][]{cebron2012letter}. Outside of the band of resonance, we find that the background flow is laminar and the zonal flow becomes independent of the geometry at first order, in agreement with a non linear mechanism in the Ekman boundary layer \cite[e.g.,][]{calkins2010lib,sauretjfm2012}.  

\end{abstract}
\end{frontmatter}
%%%%%%%%%%%%%%%%%%%%%%%%%%%%%%%%%%%%%%%%%%%%%%%%%%%%%%
\section{Introduction}
Librations, oscillatory motions of the figure axis of a planet, arise through gravitational coupling
between a quasi-rigid celestial object and the main gravitational partner about which it orbits \cite[][]{yoder95,comstock03}. Several librating bodies also possess
a liquid layer, either an iron rich liquid core like on Mercury, Io, Ganymede,
and the Earth's Moon, and/or a subsurface ocean like on Europa, Titan,
Callisto, Ganymede and Enceladus
\cite[][]{anderson96,anderson98,anderson01,williams01,spohn03,hauck04,breuer07,margot07,williams07,lorenz08,hoolst08}.
The interaction of the fluid layer with the surrounding librating
solid shell resulting from viscous, topographic, gravitational or
electromagnetic coupling leads to dissipation of energy and angular
momentum transfer that need to be accounted for in thermal history
and orbital dynamics models of these planets.

There is a whole variety of celestial objects to which our approach will be applicable, in the next paragraphs we propose to focus on the Earth's moon purely for pedagogical reasons to clarify some astronomical aspects of the problem and distinguish between the different types of librations. In Figure \ref{torque} we illustrate the origin of the gravitational torques producing the librations in the case of the Earth-Moon system considering only the principal harmonic at the orbital period. Over geological time scales, the Lunar mantle has been tidally deformed  into a triaxial ellipsoid, resulting in a mass anomaly.

Due to the eccentricity of its orbit, the Moon orbital period varies along the orbit according to the third Kepler's law. As illustrated on Figure \ref{torque}a), the induced phase lag between the Earth-Moon direction and the orientation of the equatorial bulge, the so-called {\it optical} longitudinal libration, produces a restoring torque along the spin axis of the Lunar mantle. This time periodic torque forces the Moon to {\it physically} oscillate axially about its state of mean rotation. This small oscillation is referred to as the {\it physical} longitudinal libration. Note that the optical libration is typically of the order of $0.1-1$ rad whereas the physical libration is only of order $10^{-4}$ rad due to the large inertia of the Lunar mantle.

In addition, the Moon is in a Cassini state, i.e. the relative orientations of the normal to the ecliptic plane, the spin vector of the Moon and the normal to the orbital plane of the Moon are fixed. As illustrated in Figure \ref{torque}b), it yields a gravitational torque, fixed in the frame rotating with the moon, that tends to align the equator of the Lunar mantle with the orbital plane of the Moon. The spin axis of the Moon being fixed relative to the normal to the orbital plane, the mantle oscillates about an equatorial axis perpendicular to the Earth-Moon direction resulting in the so called {\it physical} latitudinal libration. As with longitudinal libration, the {\it physical} latitudinal libration is several orders of magnitude smaller than the {\it optical} latitudinal libration. 

In contrast with precession or nutation that are well represented by gyroscopic motions of the solid shell, it is important to note that librations both in longitude and latitude do not result in changes of the orientation of the spin axis of the planet on diurnal time scales. The combination of {\it optical} librations both in longitude and latitude can be observed in an sequence of NASA images of the Moon taken from the Earth along its orbit (http://en.wikipedia.org/wiki/Libration).

The mechanical forcing produced by the two components of libration that drive the flow in the liquid layer of a planet can be illustrated by two concept laboratory experiments,  as illustrated in Figure \ref{librations}. A turntable mimics the mean rotation of the planet while the oscillation of the planet's solid shell is achieved by a mechanical system attached to the rotating table. Longitudinal libration, a time periodic oscillation of the body's figure axis about its mean rotation axis, can be simulated by oscillating the container about the vertical axis (Figure \ref{librations}a). Latitudinal libration, a time periodic oscillation of the figure axis about an equatorial axis that is fixed in the rotating frame (the turntable), is illustrated in Figure \ref{librations}b. In the present paper we only consider the flow driven by physical longitudinal libration, herein referred to as longitudinal libration.

Several of experimental, numerical and theoretical studies have
been devoted to libration-driven flows in axisymmetric containers to
investigate the role of the viscous coupling in librating planets.
It has been shown that longitudinal libration in an axisymmetric
container can drive inertial modes in the bulk of the fluid as well
as boundary layer centrifugal instabilities in the form of
Taylor-G\"ortler rolls \cite[][]{aldridgephd,
aldridge69,aldridge75,tilgner99,noir09,calkins2010lib,noir2010zf_a,sauret2012_pof}. In addition,
laboratory and numerical studies
\cite[][]{aldridgephd,wang1970,calkins2010lib,noir2010zf_a,sauret2010,sauret2012_pof} have
confirmed that non-linear interactions within the Ekman boundary layers generate a steady, axisymmetric flow, called 
zonal flow. Analytical derivations of the zonal flow driven by longitudinal libration have been carried out in cylindrical cavity for an arbitrary libration frequency \cite[][]{wang1970}, in spherical geometry at low libration frequency \cite[][]{busse2010zf_b} and more recently in spherical geometry at an arbitrary frequency \cite[][]{sauretjfm2012}. 

Although practical to isolate the effect of viscous coupling, the
spherical approximation of the core-mantle or ice shell-subsurface
ocean boundaries, herein generically called the CMB, is not physical from a
planetary point of view and very restrictive from a fluid dynamics
standpoint. Indeed, due to the rotation of the planet, to the
gravitational interactions with companion bodies and to the low
order spin-orbit resonance of the librating planets we are
considering, the general figure of the CMB must be ellipsoidal with
a polar flattening and a tidal bulge pointing on average toward the
main gravitational partner
\cite[cf.][]{Goldreich2010}. This assumes
that the libration period is far shorter than a typical deformation time scale
of the CMB.
 
In contrast with the spherical or cylindrical geometry, it has been demonstrated analytically that longitudinal libration in a non-axisymmetric spheroid can not produce resonance through direct forcing of a single inertial mode when $\epsilon^2 >> E^{1/2}$, where $\epsilon$ is the ellipticity \cite[][]{Zhang2011JFM}. This can be recast as $2\beta/(\beta+1)>> E^{1/2}$ using our notation presented in the next section, which is satisfied for the two non-axisymmetric containers considered in the present study.  

Recent analytical and numerical work by \cite{cebron2012letter} demonstrates, however, that triadic resonances are possible between two inertial modes and the elliptically deformed basic flow, leading to the so-called Libration Driven Elliptical Instability (LDEI). The elliptical instability can stably saturate in a narrow range of libration amplitude in the immediate vicinity of instability threshold \cite[][]{kerswell98,herreman2009,Cebron2011}. Outside of this limited window, a transition occurs that lead to the development of space-filling turbulence \cite[][]{malkus1989}. This turbulence, which acts to disrupt the elliptically unstable base state, decays and the flow `relaminarizes'.  The relaminarization phase ends when the base state re-establishes itself. The elliptical instability will then give way again to turbulence. In planetary liquid cores, such
an instability could be responsible for an increased
viscous dissipation \cite[][]{LeBars2010}, for the induction of a magnetic field \cite[][]{kerswell98,herreman2009}, or the presence of a dynamo \cite[][]{lebars2011_nature}. 

Finally, longitudinal libration in a non-axisymmetric ellipsoid can excite instabilities, which develop as the shell rotation is slowing down during the libration cycle \cite[][]{Liao:2011p8281}, at sufficiently low frequency and large amplitude. In contrast with the side wall centrifugal instability observed in axisymmetric container, the unstable region extends further inside the fluid interior. The underlying mechanism as well as the scaling of the threshold of these instabilities have yet to be investigated via a systematic exploration of the parameter space.	

This paper aims at describing the zonal flow driven by longitudinal librations in non-axisymmetric
ellipsoidal containers, for which we expect the topographic coupling to be dominant. Spherical and hemispherical containers have been included for comparison to emphasize the effect of the topography. 

In section \ref{math} we present the theoretical frame work for libration driven flow, the experimental method is described in section \ref{exp_setup}, section \ref{res} presents the experimental results. Finally, implications for planets and moons are considered in section \ref{conclusion}.  
\section{Mathematical background and control parameters}\label{math}
Let us consider a homogeneous, electrically non-conductive and
incompressible fluid enclosed in a librating triaxial ellipsoidal
cavity. The
equation of the ellipsoidal boundary can be written (Figure
\ref{oblong_spheroid}) 
\eqi
\frac{x^2}{a^2}+\frac{y^2}{b^2}+\frac{z^2}{{c}^2}=1,
\eqo 
where ($x,\,y,\,z$) is a cartesian coordinate system with its origin at
the center of the ellipsoid, $\vect{\hat{x}}$ is along the long
equatorial axis $a$, $\vect{\hat{y}}$ is along the short equatorial
axis $b$, and $\vect{\hat{z}}$ along the rotation axis $c$. We
define the ellipticity $\beta$ as 
\eqi
\beta=\frac{a^2-b^2}{a^2+b^2}, 
\eqo 
and the aspect ratio 
\eqi
c^*=\frac{c}{R}, \eqo where $R$ stands for the mean equatorial radius $R=\sqrt{(a^2+b^2)/2}$. In the inertial frame, the longitudinal librating motion of the container can be modeled by a time dependence of its axial rotation
rate:
\eqi \Omega(t)=\Omega_0 + \Delta\phi \, \omega_l \sin (\omega_l t).
\eqo 
Here, $\Omega_0$ represents the mean rotation rate, $\Delta\phi $ is the
amplitude of libration in radians and $\omega_l$ is the angular
frequency of libration.

To allow for an easy comparison with previous analytical work, we present the mass conservation and the momentum equations in the frame of reference attached to the librating container. Using $a$ as the length scale and
$\Omega_0^{-1}$ as the time scale, these equations are written

\begin{eqnarray}\label{NS}
\frac{\partial \vect{u}}{\partial t}-\vect{u}\times\left( \nabla \times \vect{u} \right) + 2(1+\varepsilon \sin ft)\vect{\hat{z}}\times \vect{u}=&\cr
-\nabla \pi + E\nabla^2 \vect{u} -\varepsilon f \cos ft (\vect{\hat{z}}\times\vect{r})&,\\
\nabla \cdot \vect{u}&=0.
\end{eqnarray}
The first two terms on the {left} hand side of (\ref{NS}) are the standard material derivative of the velocity field; 
the third term is the Coriolis acceleration.  The {right} hand terms are, respectively, the pressure force, the viscous force and the Poincar{\'e} force.
In (\ref{NS}), $\pi$ is the reduced pressure, which includes the time-variable
centrifugal acceleration. The Ekman number $E$ is defined by
\eqi E=\frac{\nu}{\Omega_0 a^2}, \eqo
where $\nu$ is the kinematic visosity. The dimensionless libration frequency $f$ is defined as
\eqi f=\frac{\omega_l}{\Omega_0}. \eqo
Lastly, $\varepsilon$ is the libration forcing parameter defined by
\eqi \varepsilon=\Delta\phi f. \eqo

Typical values of the dimensionless parameters for planets of our solar system are presented in table \ref{table_planets} (courtesy of \cite{noir09}).
The viscous solution to (\ref{NS}) must satisfy the no-slip boundary
condition on the CMB
\eqi \vect{u}=\vect{0} \quad \text{at}\quad %AS
x^2+\frac{1+\beta}{1-\beta}\,y^2+\frac{1+\beta}{c^2}\,z^2=1.
\label{BC} \eqo

In the limit of small Ekman number, the flow can be decomposed into
an inviscid component $\vect{U}$ in the volume and a boundary layer
flow $\tilde{\vect{u}}$. Introducing this separation, \cite{kerswell98} proposed the following solution to the inviscid equations of motion subject to the non-penetration condition at the
CMB:
\begin{eqnarray}
\vect{U}&=&-\varepsilon \sin ft \left( \vect{\hat{z}}\times\vect{r}-\beta\nabla xy \right),\label{sol_inv1}\\
\pi&=&-\varepsilon f \beta xy \cos ft + \varepsilon \sin ft \left(1+
\sin ft \right) \left( \left | \vect{\hat{z}}\times \vect{r}\right
|^2 +\beta \left( x^2-y^2 \right)  \right) . \label{sol_inv2}
\end{eqnarray}
The base flow $\vect{U}$ is the sum of a time dependent uniform vorticity flow and a gradient component. It follows that the Reynolds stresses resulting from (\ref{sol_inv1}) are balanced by
the pressure gradient. Therefore, no net zonal flow can result from
the non-linear interactions in the quasi-inviscid interior
\cite[][]{busse2010zf_b}. However, the no-slip boundary condition
(\ref{BC}) is not entirely fulfilled by this inviscid solution. Hence, viscous
corrections in the Ekman boundary layer must also be considered. Their non-linear interactions can generate zonal flow in the bulk \cite[][]{wang1970,busse2010zf_a,busse2010zf_b}, as
already observed in spherical and cylindrical (i.e., axisymmetric) geometries
\cite[][]{aldridgephd,wang1970,calkins2010lib,noir2010zf_a,sauret2010,sauret2012_pof}.
In the present paper we use the analytical derivation of the zonal flow from \citet{sauretjfm2012}, an outline of the method is presented in the Appendix.

\section{Experimental method.}\label{exp_setup}
Figure \ref{Exp_device} represents a schematic view of the
experimental device used in the present study. Except for the
containers, the laboratory apparatus is the same as in \cite{noir09} (see section 3.1 for a detailed description). The generic set-up consists of a turntable
rotating at a constant angular velocity $\Omega_0$ and an oscillating acrylic tank centered on the turntable activated by a brushless direct drive motor. Both rotations are controlled using a motion control system that allows for high accuracy, better than $0.1\%$ on the mean rotation and $0.25\%$ on the angular displacement. The container consists of two ``hemispheres'' CNC machined from
cast acrylic cylindrical blocks that are polished optically clear. To characterize the effect of the topographic coupling resulting {from} the non axisymmetry of the librating body, we use three different {containers}: i) a sphere of radius
$a=127$ mm, ii) a prolate {spheroid} of long axis $a=127$ mm
and short axis $b=c=119$ mm and iii) a prolate {spheroid} of long axis $a=127$ mm
and short axis $b=c=89$ mm. These containers correspond to an ellipticity in the equatorial cross section
$\beta$ equal to 0, 0.06 and 0.34 respectively. In all three {configurations} the rotation axis is along c.

We perform direct visualizations
of the interior flows using a diluted solution of rheological fluid (Kalliroscope), and a
horizontal or vertical laser light sheet. A CCD camera located above the container
records movies and still images to characterize the time evolution
of the shear structures in the interior. In addition,
we use a remotely controlled syringe pump to inject dye (fluorescein
or non-diluted Kalliroscope) at a cylindrical radius $s_i\sim0.38$
along the short axis of the mean elliptical equator, i.e. the time averaged figure axis of the
equatorial cross-section (see Figure \ref{Tank_geometry}c). We then manually track the dye over a full
revolution (until it passes by the injection point again) or a
fraction of a revolution when the patch spatial coherence is lost
due to turbulent mixing. These observations are used to derived the
mean angular velocity along the elliptical path followed by the dye
as illustrated in Figure \ref{Tank_geometry}c. Error bars are obtained by repeating the dye injection several time during the same experiment. Although {straightforward}, this technique is not suitable when the azimuthal velocity varies on time scale less than a period of revolution of the dye patch.

To address the time dependency of the azimuthal velocity in the system,
we performed LDA measurements using the ultraLDA system employed in \cite{noir2010zf_a}. The point of measurement as been choose as to coincide with the dye inlet position at times $t=0+N/2f$. For detailed description of LDA
principles and measurement techniques we refer the reader to the
appendix A of \cite{noir2010zf_a}. 
In principle an LDA device works
as follows: a laser beam is split to produce two beams that are
collimated at a point inside the liquid where it forms a linear
pattern of interference fringes. Particles in suspension in the fluid act
as reflectors when passing through the fringe pattern resulting in back scattered light that is focussed on a
photodetector. A spectral analysis of the received signal leads to a
measurement of the velocity in the direction perpendicular to
fringes.

Due to the difference of index of refraction of water, acrylic and air, the laser beams traveling through the system experience two optical distortions at the
air-acrylic and water-acrylic interface. The
laser beams will be deflected both in longitude and latitude independently of one-another depending
on their orientation with the local normal to the surface. In some situations, the two beams may no longer be coplanar, precluding any measurements. This is indeed the case in all possible configurations with the current experimental setup. In order to overcome this limitation, we perform LDA measurement with the northern half of the
container replaced with a flat plate of acrylic. The LDA device
is located above the tank and oriented to perform measurements of
the azimuthal component of velocity (Figure \ref{Exp_device}). Such geometry retains the non axisymmetry of the equatorial cross section and the latitudinal variation of the wall curvature. As it will be shown in the next
section, the mean zonal flows generated by the libration of the full
and half containers are in quantitative agreement over a broad range
of parameters. However, it must be noted
that only resonant modes where the vertical velocity at the
equator is zero can be excited in the hemispherical configurations. One may envisage further implications when substituting the upper hemi-sphere/ellipsoid with a flat lid such as viscous drag, equatorial edges driven flows or reduced effects of curvature. As we we shall see in the following of this paper, quantities like the zonal flow does not significantly differs between full and hal container. This is indeed supported by the similar analytical prediction obtained in spherical and cylindrical geometries by \cite{busse2010zf_a} and \cite{sauretjfm2012}.   

In order to get the best signal-to-noise ratio, we use a
fresh suspension of titanium micro-particles, $TiO_2$, every day. The data rate of LDA
measurements varies typically between 25Hz and 500Hz. In order to
perform spectral analysis and proper time averaging, each time series
is resampled at 10Hz using the non-linear interpolation routine of Matlab to
provide an equally spaced dataset. The mean zonal flow is obtained
by block averaging the data, each block is 20 libration periods wide, each record is 200 periods long in the steady state (i.e. after several spinup times). The error bars {represent} the variability of the zonal flow from block to block. We obtain error bars of the order of $2-5\%$ at moderate forcing ($\varepsilon<1.5$) and $15-20\%$ for $\varepsilon>1.5$. Finally, when laminar-turbulence
intermittency is observed, we perform moving average over a window of 10
periods of libration with an overlap of 90$\%$ to characterize the zonal flow in each phase of the system.

\begin{flushleft}
\begin{table}[h]
\begin{center}
\begin{tabular}{|c|c|c|}
\hline
Parameter & Definition& Experiment \\
\hline
$a$ & long axis & 127 mm\\
$b$ & short axis & 89 mm, 119 mm and 127 mm\\  
$c$ & short axis & 89 mm, 119 mm and 127 mm\\  
$S_i$& Injection point radius & 48mm\\
$\Omega_0/2\pi$ &  Mean rotation frequency& 0.5 Hz\\
$\omega_L/2\pi$&Libration frequency& 0.25 - 1 Hz $\pm 0.1\%$\\
$\Delta \phi$&Angular displacement & 0 - $\pi/2$ \\
$\nu$ & Kinematic viscosity & $10^{-6}$ m$^2$s$^{-1}$ \\
\hline
$E$&$\nu/(\Omega_0\;a^2)$& $2.0\times10^{-5}$ \\
$f$&$\omega_{L}/\Omega_0$&0.5 - 2 \\
$\beta$&$\frac{a^2-b^2}{a^2+b^2}$&0.34, 0.06 and 0\\
$c^*$&$\frac{c}{\sqrt{(a^2+b^2)/2}}$&0.812, 0.967 and 1\\
$\varepsilon$ & $(\Delta\phi)\,f$&0 - 1.6\\
$s_i$ & $S_i/a$ & 0.38\\
\hline
\end{tabular}
\caption{ Physical and dimensionless parameters definitions and
their typical values in the laboratory experiment, with the rotation axis $c=b$. }
\end{center}
\end{table}
\end{flushleft}

Each experiment follows a common protocol. First, we start the
rotation of the turntable, once the fluid {is} in solid body rotation we
turn on the oscillation of the container. When performing dye
tracking, we wait 15 min before injecting the dye and recording from
the CCD. When performing LDA measurement, we start the acquisition as we turn on the libration
to follow the development of the dynamics until it reaches a steady
state. 

The physical and dimensionless
parameters accessible with the present device are summarized in
{table 1}. In contrast with the previous studies using this device, we
fix here the mean angular velocity to $\Omega_0=30$ rpm,
corresponding to an Ekman number $E=2\times 10^{-5}$, and we explore
the parameter space $(f,\Delta\phi)$. 
\section{Results}\label{res}

In Figure \ref{ZF_f05hz_various_dphi}, we present the time averaged zonal
flow from direct visualization and LDA measurements together for a
fixed Ekman number, $E=2\times10^{-5}$, a fixed libration frequency,
$f=1$, and a libration amplitude $\Delta\phi$ that varies from 0.05
to 1.6 rad. In all cases, the mean zonal flow is retrograde at the point
of measurement. In the range of studied parameters and at the measurement location $s_i=0.38$, we do not observe significant %AS
variations of the zonal flow amplitude with the ellipticity, nor
with the half or full tank configurations.  Hence, we expect the same mechanism, weakly dependent on the geometry, to produce the zonal flow
in all 6 configurations (full tank, $\beta=0,0.06,0.34$; half tank $\beta=0,0.06,0.34$).
\cite{busse2010zf_b,busse2010zf_a,calkins2010lib,noir2010zf_a,sauret2010,sauretjfm2012} have
proposed that the geostrophic zonal flow in librating axisymmetric containers
results from boundary layer non-linear interactions, which yields a zonal flow independent of the Ekman number at the first order, scaling as $\zf \propto \varepsilon^2$. \cite{sauretjfm2012} recover the pre-factor and the radial dependency of the geostrophic flow by deriving the boundary layer flow at the order $\mathcal{O}(\varepsilon^2E^{1/2})$ in a full sphere (see Appendix). For a probe volume located at $s_i=0.38$, same as dye injection location, and a frequency $f=1$, the authors predicts a zonal flow, $\zf =\alpha \varepsilon^2$, with $\alpha=-0.166$. We observe a good agreement between their analytical spherical model represented by the dashed line in Figure \ref{ZF_f05hz_various_dphi} and our LDA measurements in all 6 configurations up to $\Delta\phi=1.6$ rad. The theoretical zonal flow in a non-axisymmetric container has yet to be derived. Nevertheless, our results suggest that the full sphere model remains a good approximation at first order even for finite ellipticity, highlighting the minor role of the curvature in the source mechanism. 

Significant deviations only appear for the largest value of $\Delta\Phi$ studied here, where we observe a larger zonal
flow than predicted by the non-linear analysis valid only for $\varepsilon << 1$. It is likely that for $\varepsilon \gtrsim 1$,
the boundary layer flow derivation of \cite{sauretjfm2012} is not valid anymore and finite amplitude
effects should be introduced.

In Figure \ref{ZF_dphi80_various_f}, we present LDA measurements of the
time average zonal velocity as a function of the libration frequency at
$E=2\times10^{-5}$ and a fixed libration amplitude $\Delta\phi=0.7$
rad. All measurements are performed only in the half container
geometry using LDA to diagnose the flow. We also plot the analytical geostrophic zonal velocity for a probe volume located at $s_i=0.38$ derived from \cite{sauretjfm2012} for a full sphere (see Appendix). 

For $\beta=0$ and $\beta=0.06$, we observe only marginal differences
between the different half containers up to $f\sim1.6$: the flow remains laminar and the experimental results are consistent with the
retrograde geostrophic zonal flow predicted from the non-linear boundary layer mechanism. At higher frequencies, the theoretical analysis of \cite{sauretjfm2012} predicts a geostrophic discontinuity in the zonal flow associated with the so-called critical latitude. As we scan in frequency, the geostrophic shear structure passes by the measurement point when $s_c=s_i=0.38$, i.e. $f=1.85$. In Figure \ref{ZF_dphi80_various_f}, our LDA measurements do not show a sharp transition in this frequency range and we note a significant discrepancy between the predicted zonal flow and the time averaged velocity measurements. Several effects can account for this disagreement. First, in the theoretical analysis, the Ekman boundary layer becomes singular at the critical latitude resulting in a local infinite geostrophic shear. This discontinuity can be resolved by taking into account higher order terms. Doing so, the geostrophic shear has  a finite amplitude and occurs in a $E^{1/5}$ width layer centered on the critical cylindrical radius $s_c$ \cite[e.g.][]{noir01num, kida2011}. Thus, we do not expect the analytical zonal flow derived by \cite{sauretjfm2012} to be valid when $s_c - E^{1/5}<s_i<s_c + E^{1/5}$. As we scan in frequencies, the region of influence of the geostrophic shear stucture passes by the point of measurement located at a cylindrical radius $s_i$ when $1.73<f<1.92$ (Figure \ref{ZF_dphi80_various_f}). Second, in this range of parameters $\varepsilon$ becomes significantly larger than unity. Thus, we expect finite amplitude perturbations, not taken into account in the analytical model, to contribute to the local mean zonal velocity.

The case $\beta=0.34$ is more complex. Indeed, for $f\in [1.43; 1.66]$, we observe intermittency of lower and higher amplitude zonal flows represented by open diamonds and red full diamonds, respectivey. This intermittency is illustrated in Figure \ref{ZF_30RPM_073HZ_80DEG_LDA_VIDEO}, which represents the time evolution of
the norm of the azimuthal velocity averaged over 10 oscillations for a particular experiment at $\Delta\phi=0.7$ rad ($\varepsilon
\sim1$), $f=1.46$, $\beta=0.34$ (red) and $\beta=0.06$ (blue). At $\beta=0.34$, the zonal flow averaged over 10 oscillations evolves in time between a low amplitude $|<\vect{U_{\phi}}_1>|\sim0.09$ and a large amplitude $|<\vect{U_{\phi}}_1>|\sim0.14$. Using a diluted Kalliroscope suspension and a camera at the top with a horizontal laser light sheet 1cm below the flat top lid we visualize the shear structures in the interior. Periods of large amplitude mean zonal flow are systematically correlated with small scale shear structures, corresponding to the bright filaments in the second snapshot, whereas periods of low amplitude mean zonal flow are associated with laminar flows, which are characterized by little contrast variation as on the first and last snapshots. The duration of each laminar and turbulent period varies over the experiment but is always of the order of a fraction of the spinup time (Each black and white rectangle at the top of Figure \ref{ZF_30RPM_073HZ_80DEG_LDA_VIDEO}a represents a spinup time ($\sim 70$s) ). The transition from low amplitude to large amplitude mean zonal flow occurs over a typical timescale $\tau\sim 25$s ($\sim 10$ rotations) that remains consistent over the whole experiment. This particular dataset is representative of all experiments where intermittent turbulence is observed. The typical time scales of the herein reported intermittent turbulence are not consistent with the centrifugal instability observed in the spherical shell \cite[][]{noir09} or in previous numerical simulations in non-axisymmetric ellipsoids at low libration frequencies by \cite{Liao:2011p8281}, which both occur once per libration period. 

In the band of frequency $f\in [1.43; 1.66]$, the zonal flow predicted by \cite{sauretjfm2012} is significantly different from our observations, in particular it fails at reproducing the intermittent low and large amplitude zonal flow.

\section{Discussion and concluding remarks}\label{conclusion}

In the present study we explore the zonal flow regimes driven by longitudinal libration in the ($f,\,\,\Delta\phi$)-parameter space at $E=2\times 10^{-5}$ for spherical and non-axisymmetric containers. At fixed frequency $f=1$ the flow in the bulk remains laminar for all accessible amplitudes of libration $\Delta \Phi$ regardless of the tank geometry. In this laminar regime, we measure a net zonal flow that is independent of the geometry at first order and well explained by non-linearities in the Ekman boundary layer.  In contrast, at a fixed amplitude of libration $\Delta \Phi=0.7$ rad, we observe space-filling turbulence correlated with an enhanced zonal flow in specific bands of frequency and for the container with the largest equatorial ellipticity. Using two containers {with different equatorial ellipticities} and a spherical cavity, we unambiguously demonstrate that the observed instability results from the topographic coupling and not from viscously-driven dynamics. Although the range of accessible parameters in our device does not allow us to study in great details the mechanism underlying the onset of the turbulent regimes, some possible routes can be investigated.  

Comparing Figure \ref{ZF_f05hz_various_dphi} and Figure \ref{ZF_dphi80_various_f} for
$\varepsilon > 1$, our results suggest that the onset of instability is not characterized by a critical Rossby number, $\varepsilon_c$. For $f=1$ no turbulence is observed in any of our {containers} even at
the largest libration amplitude accessible in this experiment, which
corresponds to a Rossby number $\varepsilon\sim1.6$. In contrast, the intermittent turbulence is observed in the container of large ellipticity in a band of frequency $f \in
[1.43; 1.66]$, corresponding to $\varepsilon \in
[1; 1.16]$. This highlights the peculiar role of the ellipticity and frequency in the destabilization mechanism in the system.     

In rapidly rotating non-axisymmetric container, intermittency of turbulent flows followed by a re-laminarization in particular bands of frequency are often typical of the growth and collapse of an elliptical instability \citep{malkus1989}. Furthermore, recent numerical and theoretical work by \cite{cebron2012letter} has demonstrated that longitudinal libration can drive elliptical instability in triaxial and biaxial ellipsoids. In order to test if such a mechanism could explain our observations we calculate the growth rate predicted by the geometry-independent WKB analysis \cite[][]{cebron2012letter}:
\eqi 
\sigma = \sqrt{\sigma_{inviscid}^2 - (f_{res}-f)^2} - K E^{1/2}, \label{sigmath}
\eqo 
with
\eqi \sigma_{inviscid} = \frac{16+f_{res}^2}{64}\beta \varepsilon, \eqo
where $f_{res}$ is resonant frequency and K is a viscous dissipation factor typically in the range $[1-10]$.
Assuming the base flow (\ref {sol_inv2}) is realized in the experiment and a perfect triadic resonance at $f=f_{res}=1.5$, we obtain a negative growth rate for $\beta=0.06$ and a positive growth rate for $\beta=0.34$. In Figure \ref{ZF_30RPM_073HZ_80DEG_LDA_VIDEO} we superimposed the theoretical growth of the azimuthal velocity for $\beta=0.34$ and the two extreme values of the dissipation factor, $K=1$ (dotted black) and $K=10$ (dashed black) to the time series of LDA azimuthal flow in the half spheroid configuration. The good agreement for $K=10$, the most dissipative case, {suggests} that an LDEI mechanism may explain our observations. Note that the WKB approach is based on a local plane waves decomposition of the velocity field independent of the geometry of the container. It is therefore applicable to both the half and full container providing the same base flow is excited. The WKB analysis provides an upper bound of the growth rate. A more accurate prediction can be obtain via a global modes analysis, which is out of the scope of the present paper. In the present experiment, the limited quantitative diagnostics in this complex geometry do not allow us to draw a hard conclusion. Further numerical and experimental investigations at lower Ekman numbers will be necessary to characterize in details the mechanism underlying the intermittent turbulence and how this turbulence modifies the mean zonal flow. Exploring the parameter space using 3D numerical simulations in non-axisymmetric containers will remain limited to $E\geq 10^{-5}$. We are currently developing a new experimental setup to overcome this limitation.

At planetary settings two scenarios may be drawn. In the first scenario, the conditions required to drive intermittent turbulence are not met and the topographic coupling does not significantly alter the dynamics driven by viscous interactions in the boundary layer. In that case we expect the flow in the interior to remain laminar with a time averaged zonal component independent of the Ekman number following a quadratic scaling in the amplitude of libration. This would lead insignificant zonal flows in the range $10^{-9}$m/day $\lesssim U \lesssim10^{-5}$m/day for the celestial objects presented in table \ref{table_planets}. As proposed by \citet{calkins2010lib} such dynamics will not result in significant energy dissipation nor magnetic field generation. In the second scenario, topographically driven space-filling turbulence develop in the liquid layer of the planet. In that case one may expect significant energy dissipation and maybe magnetic field induction depending on the strength of the turbulence. 

Understanding the underlying mechanism for the instability reported in the present study is therefore fundamental for planetary applications. In this study, we suggest that an LDEI mechanism, identified numerically by \cite{cebron2012letter} in biaxial and triaxial librating ellipsoids, may be responsible for the observed space-filling turbulence at moderate Rossby numbers in our experiment.  

\section*{Appendix: Analytical determination of the mean zonal flow}

In this section, we present the main steps of the analytical derivation of the mean zonal flow induced by longitudinal libration in spherical geometry and we refer the reader to \cite{sauretjfm2012} for a more complete and generic description.   

In the limit of small Ekman number $E \ll 1$, the flow can be classically separated in two components: an inviscid component in the bulk and a viscous component in the Ekman boundary layer of size $E^{1/2}$ attached to the mantle. Using a perturbative approach in the limit of small libration amplitude $\Delta\phi \ll 1$, the flow can be written in the bulk:
\begin{equation}
\mathbf{U}=\mathbf{U_0}+(\Delta\phi\,f)\,\mathbf{U_1}+(\Delta\phi\,f)^2\,\mathbf{U_2}+o((\Delta\phi\,f)^3),
\end{equation}
\noindent and in the boundary layer
\begin{equation}
\mathbf{u}=\mathbf{u_0}+(\Delta\phi\,f)\,\mathbf{u_1}+(\Delta\phi\,f)^2\,\mathbf{u_2} +o((\Delta\phi\,f)^3).
\end{equation}
\noindent In the absence of librational forcing, the fluid is in solid-body rotation $\mathbf{U_0}=s\,\Omega_0\,\mathbf{e_\phi}$.
Then, as long as the libration period $1/f$ remains small compare to the spin-up time, $\sqrt{E}\ll f$, no spinup takes place in the bulk at each libration cycle \cite[][]{busse2010zf_b} and the first order correction of the bulk flow is null: $\mathbf{U_1}=\mathbf{0}$. However, to adjust the velocity field between the bulk and the librating mantle, a flow $\mathbf{u_1}$ oscillating at frequency $f$ develops in the thin Ekman layer. The nonlinear self-interactions of this oscillating flow lead to a nonlinear steady flow in the boundary layer at order $(\Delta\phi\,f)^2$, $\mathbf{u_2}$. The continuity of the velocity at the interface between the inviscid interior and the boundary layer implies a correction in the bulk flow at order $(\Delta\phi\,f)^2$, which generically writes $\mathbf{U_2}=s\,\Omega_2(s)\,\mathbf{e_\phi}$. The expression of $\Omega_2(s)$ depends on the libration frequency $f$ and the specific shape of the container. The solution differs when considering a flat top boundary as in the cylindrical geometry \cite[][]{wang1970} and a curved boundary as in the spherical geometry \cite[][]{sauretjfm2012}. 

\begin{center}
\begin{figure}
\begin{center}\includegraphics[width=13cm]{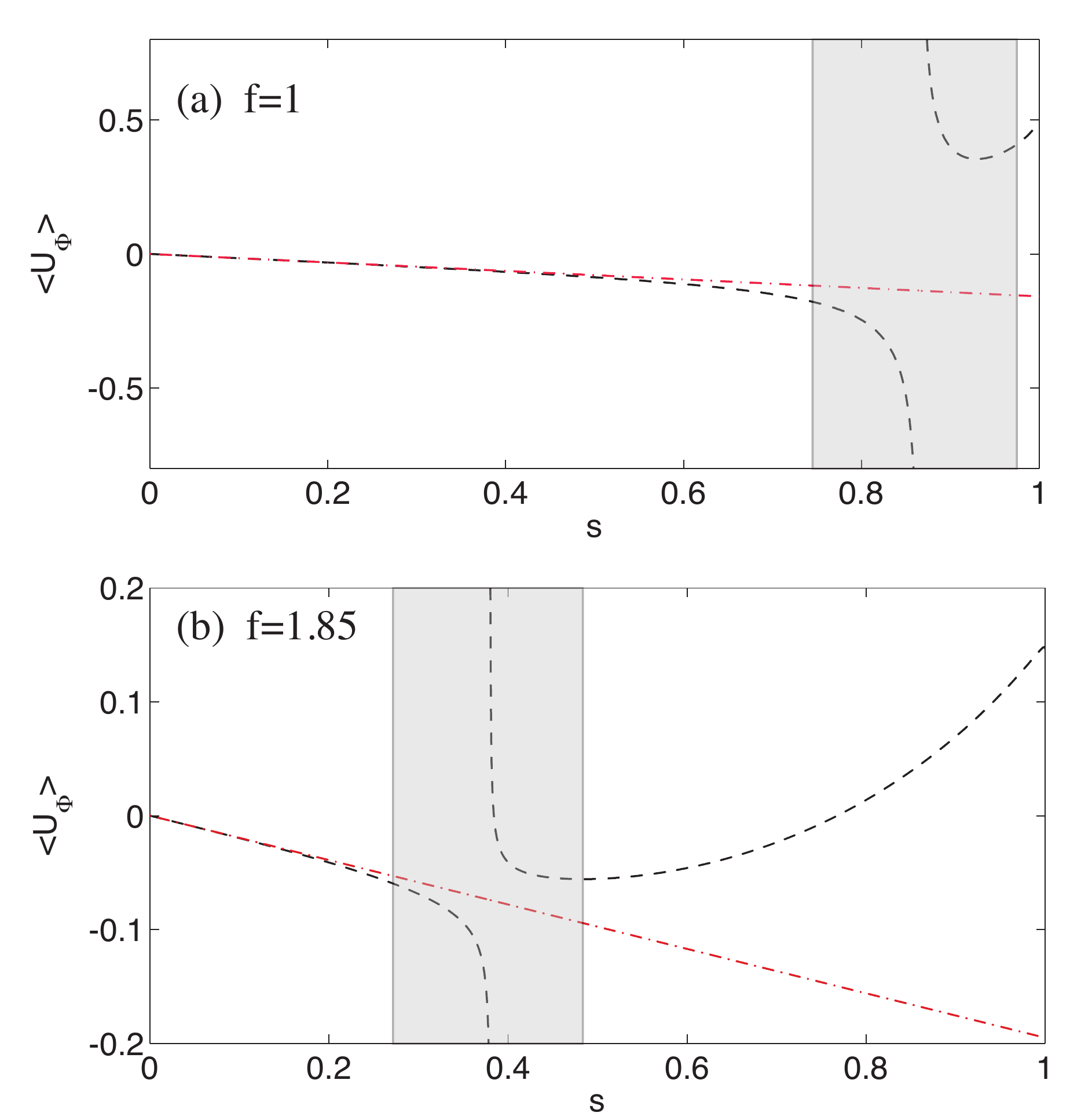}
\end{center}
\caption{Mean zonal flow $<U_\phi>$ as a function of the cylindrical radius $s$. The dashed black lines represent the analytical solution in the sphere from \cite{sauretjfm2012} and the red dashed-dotted line represents the analytical solution in a cylinder from \cite{wang1970}. (a) for $f=1$, (b) for f=1.85. The two shaded rectangles indicate the cylindrico-radial extension of the geostrophic cylinder spawn by the critical latitude in the sphere.}
\label{appendix:theory}
\end{figure}
\end{center}

In figure \ref{appendix:theory}, we show the resulting mean zonal flow $<U_\phi>=s\,\Omega_2(s)$ as a function of the cylindrical radius for two libration frequencies $f=1$ and $f=1.85$, in the case of a sphere and a cylinder. In all cases, we predict a scaling of the mean zonal flow with $(\Delta\phi\,f)^2$ but the radial profiles are different in each geometry. 

In the case of the sphere the asymptotic derivation predicts a divergent zonal flow at a critical cylindrical radius $s_c$, which corresponds to the so-called critical latitude $\theta_c$ defined as \cite[][]{lyttleton53}:
\begin{equation}
\theta_c=\text{acos}\,\left(\frac{f}{2}\right).
\end{equation}

\begin{equation}
s_c=\sqrt{1-\frac{f^2}{4}}.
\end{equation}

This critical latitude is associated with a breakdown of the Ekman boundary layer due to the total absorption of the inertial waves at this location. In the analysis of \cite{sauretjfm2012} this breakdown appears as a singularity,. However, including terms of order $\mathcal{O}(E^{1/5})$ leads to a finite amplitude shear scaling as $E^{1/5}$ over a radial and latitudinal extension scaling as $E^{2/5}$ and $E^{1/5}$, respectively \cite[][]{stewartson63,kida2011}. Using these scalings in the context of a precessional forcing, \cite{noir01num,kida2011} have proposed that the geostrophic cylinder spawn by the critical latitude scales as $E^{1/5}$ in width and $E^{-3/10}$ in amplitude. The breakdown of the Ekman boundary layer at the critical latitude and the subsequent scalings are generic to any oscillatory mechanical forcing through the boundary and remains therefore valid in the case of longitudinal libration \cite[][]{calkins2010lib}.
The complete derivation of the zonal flow including the higher order terms near the critical latitude is very fastidious and was beyond the scope of the analysis of \cite{sauretjfm2012}. Hence, we do not expect the theoretical profile derived from their analysis to apply in the range $s_c - E^{1/5}<s_i<s_c + E^{1/5}$. At a fixed measurement point $s_i=0.38$, when scanning in frequencies, the geostrophic shear alters the zonal flow measurements in a range $1.73<f<1.92$ (see Figure \ref{ZF_dphi80_various_f} in section \ref{res}).

\section*{Acknowledgements}
The authors would like to thank  F. H. Busse, S. Le Diz\`es, N. Rambaux, T. Van Hoolst and K. Zhang for
fruitful discussions on the problem. This work was financially
supported by NASA's PG\&G program (award \#NNX09AE96G), PME
program (award \#NNX07AK44G) and ERC grant (247303 MFECE).
 %%%%%%%%%%%%%BIBLIO%%%%%%%%%%%%%%%%%

\clearpage
\pict[15cm]{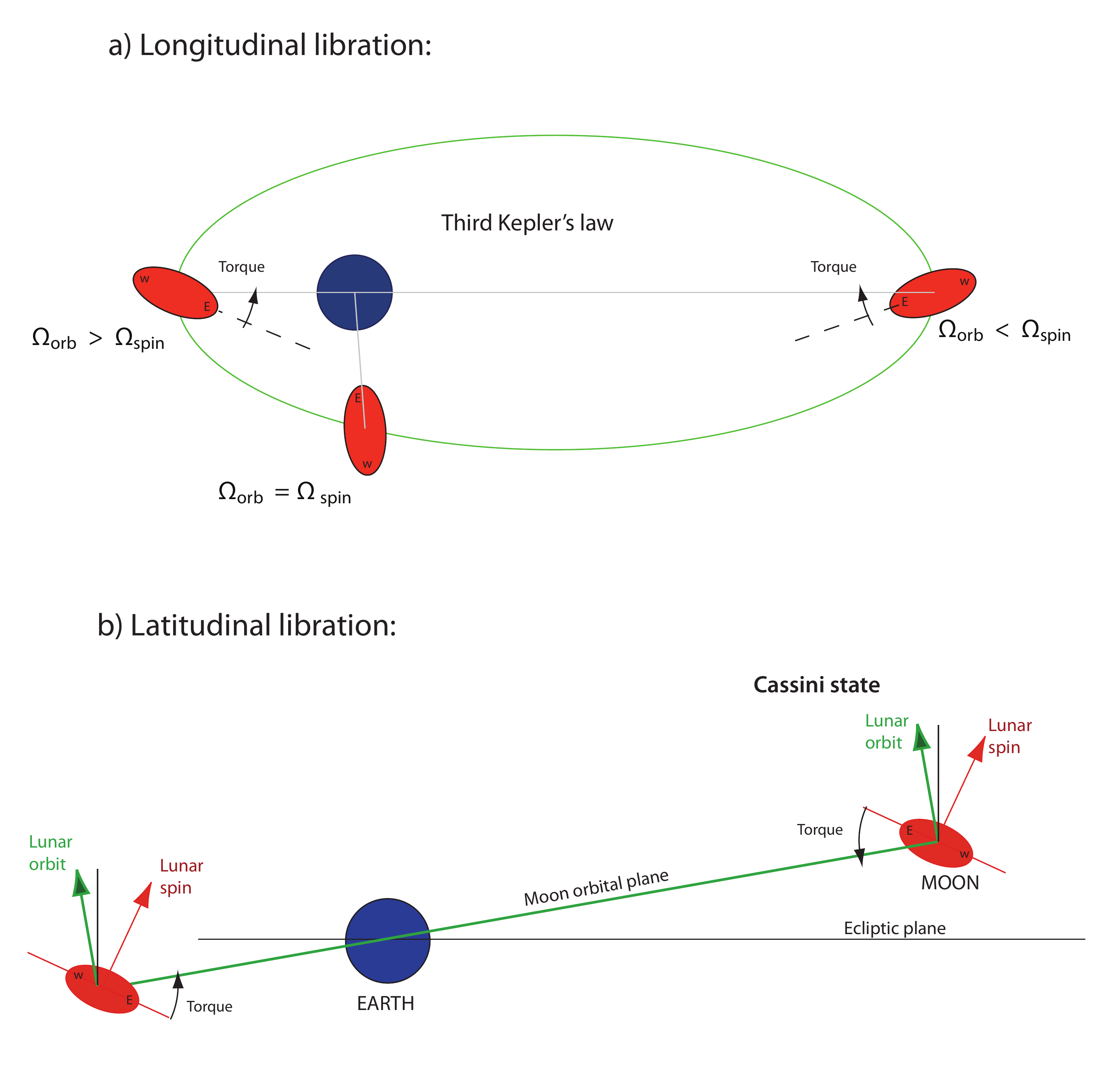}{Schematic representation of the Earth-Moon system illustrating the origin of the torque producing a) libration in longitude, b) libration in latitude. All angles have been exaggerated for clarity purposes. E and W represent two fixed point of the lunar mantle.}{torque}

\clearpage
\pict[15cm]{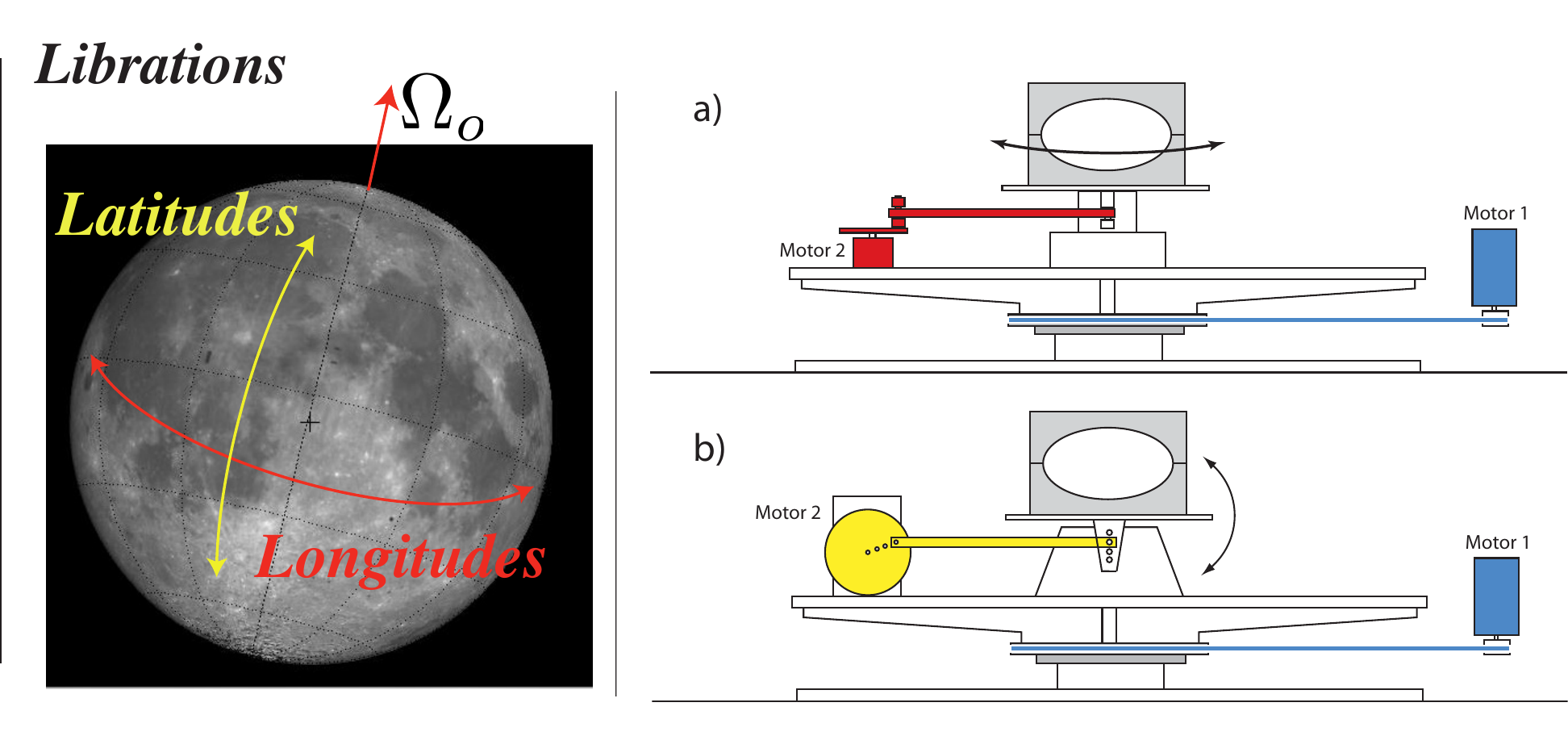}{Schematic representation of the librations in latitude (yellow) and longitude (red). The left panel represents the two {librations} for the Earth Moon, the right panel {represents} two simple schematics of experimental {setups} that mimic libration in longitude (a) and latitude (b). Note that in both cases the driving mechanism is installed on the rotating table.}{librations}

\clearpage

\pict[8cm]{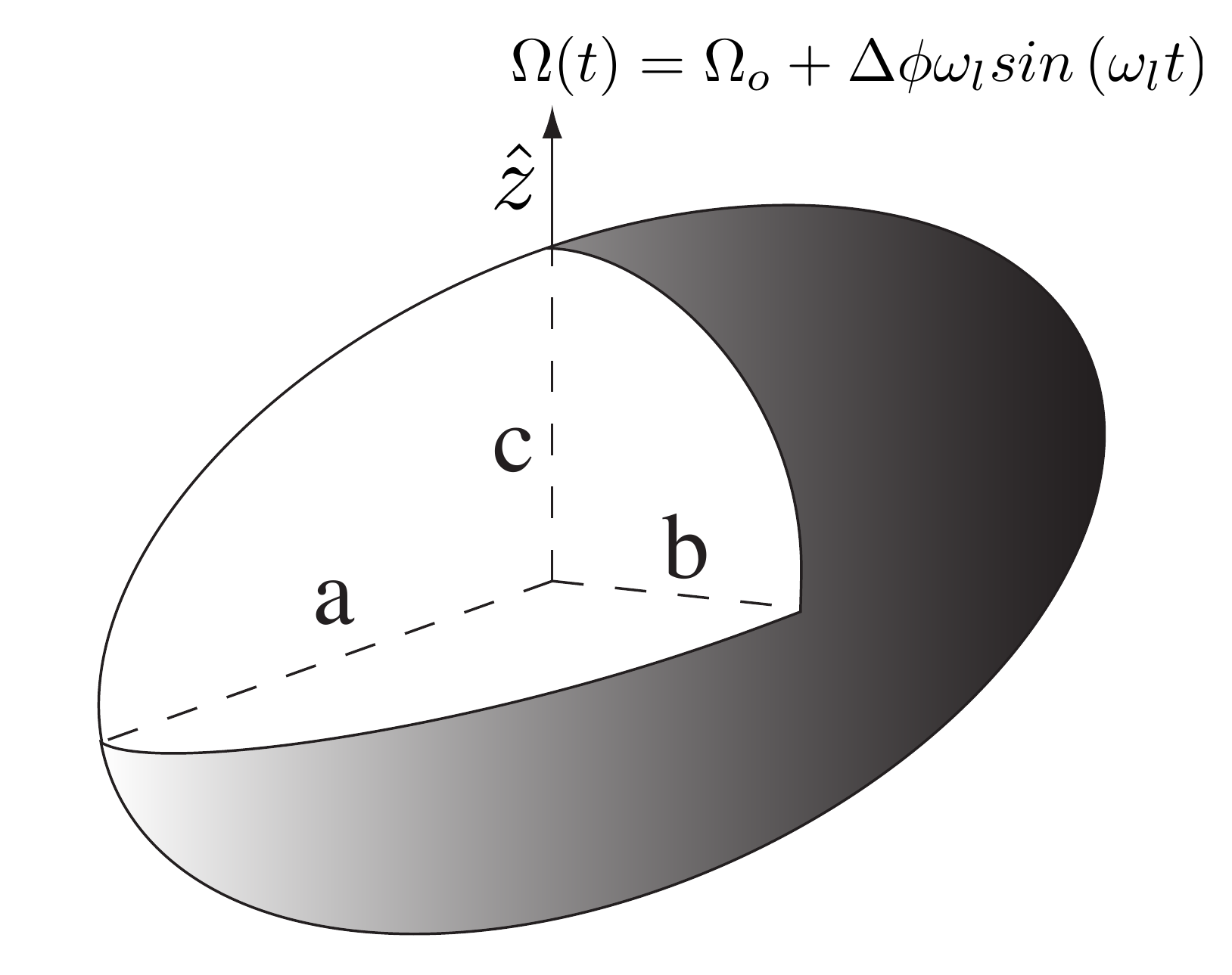}{Schematic view of the
triaxial ellipsoid.}{oblong_spheroid}

\clearpage
\pict[15cm]{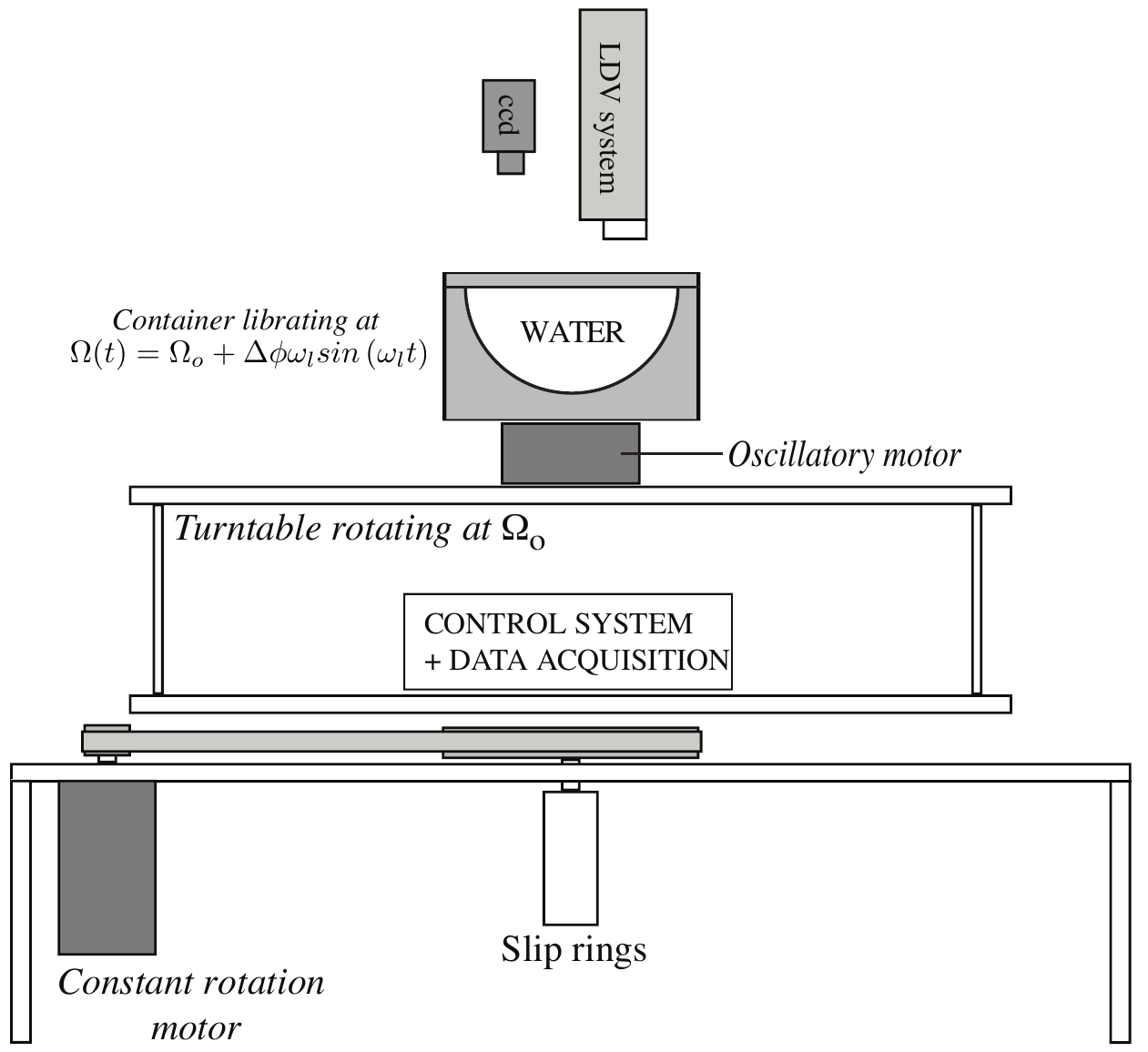}{Schematic view of the laboratory experiment, set up to acquire LDV measurements in a hemisphere or hemispheroid.}{Exp_device}

\clearpage
\pict[15cm]{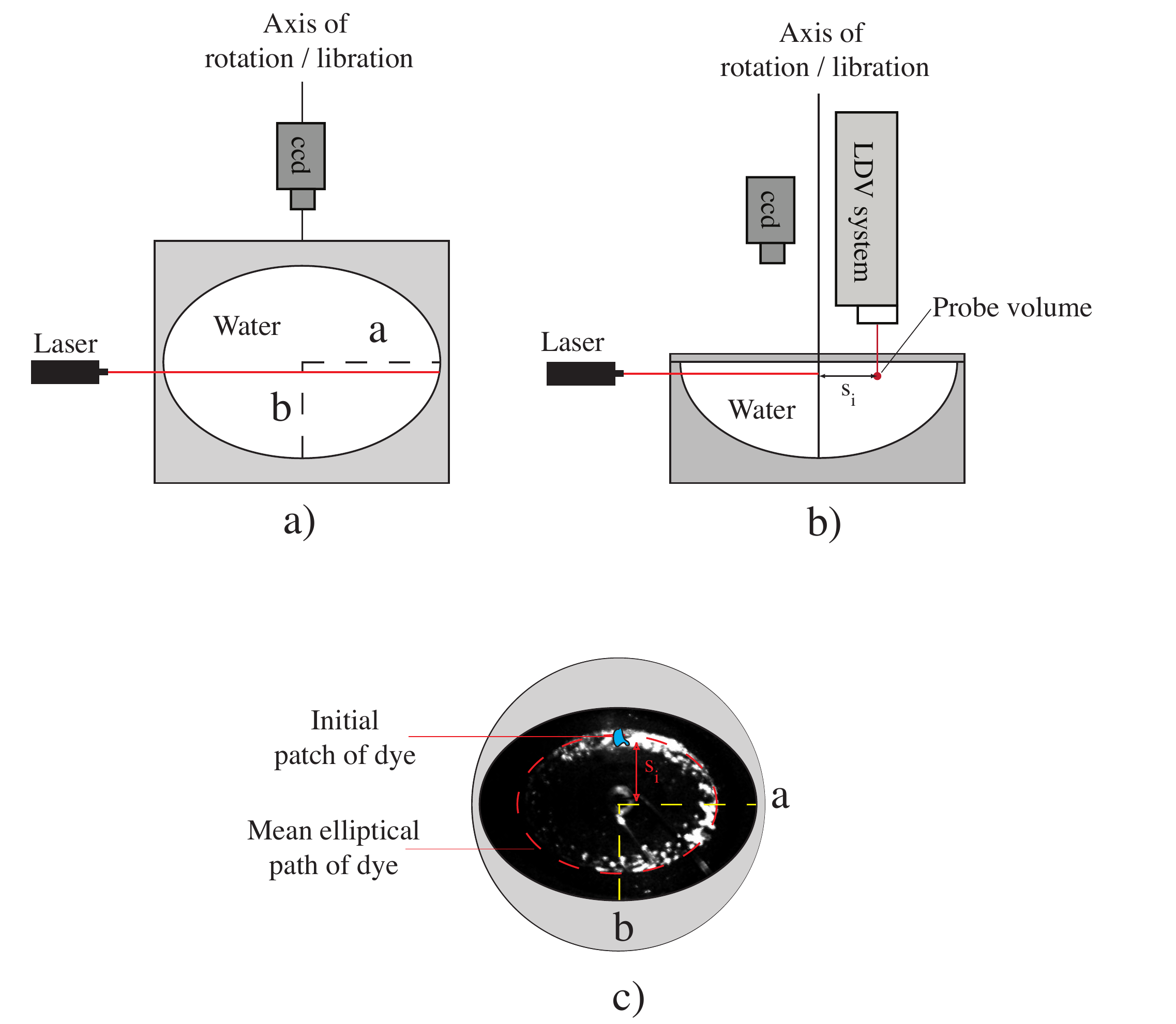}{Schematic view of the container. a)
Side view of the full container geometry. b) Side view of the half
container geometry. c) Top view for both the full and half container
geometries. The dashed ellipsoid represents the mean path of the dye
initially injected along the short axis of the container. The
picture in the interior shows the mean path using a continuous
injection of dye. The dark zone on the left results from the tilt of
the laser light sheet due to the meridional curvature of the
container.}{Tank_geometry} 

\clearpage
\pict[15cm]{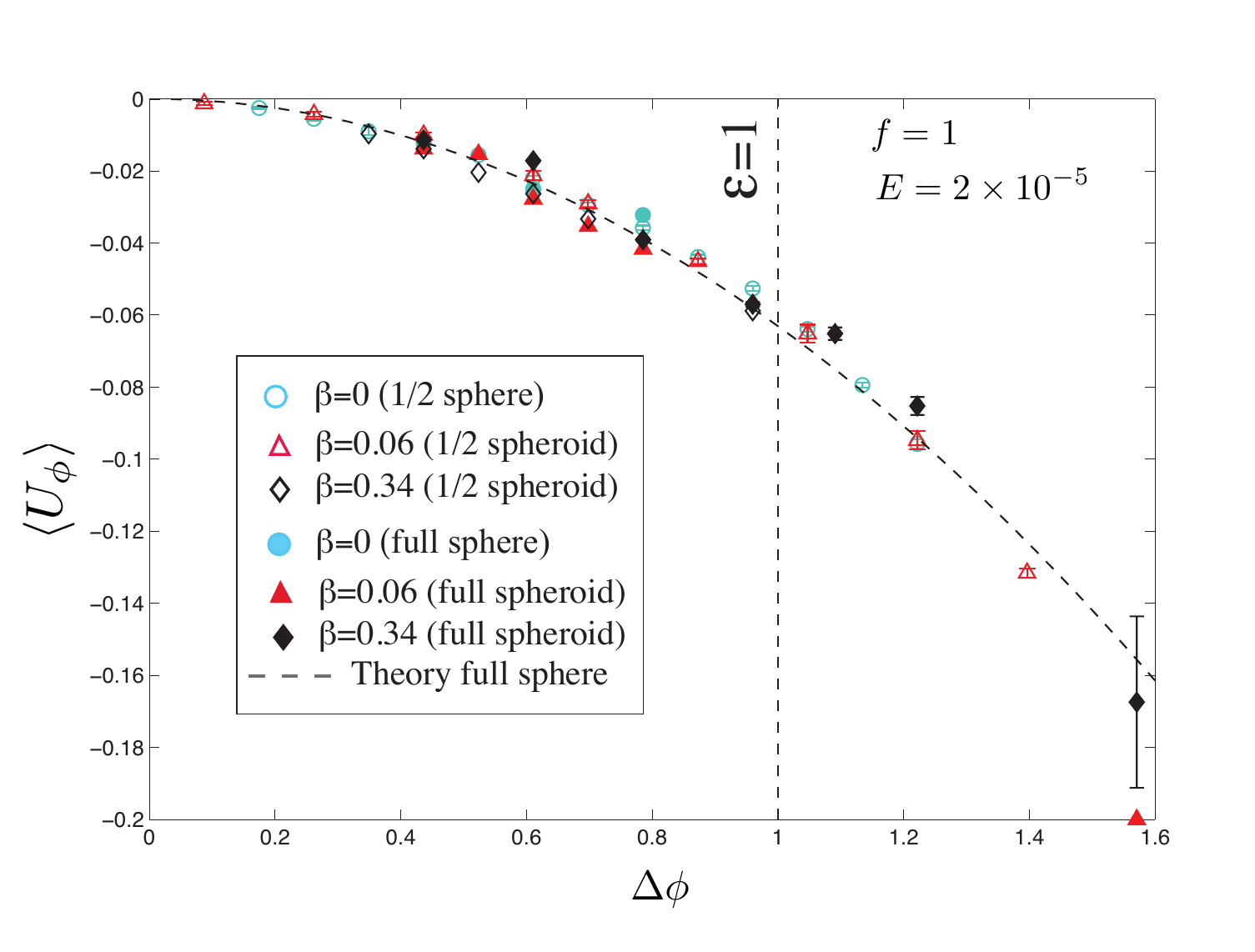}{Mean zonal flow as a function
of the libration amplitude in radians for six geometries; hemisphere
(open blue circle), full sphere (filled blue circle), 1/2 spheroid
with $\beta=0.06$ (open red triangle up), full spheroid with
$\beta=0.06$ (filled red triangle up), 1/2 spheroid with
$\beta=0.34$ (open black diamonds), full spheroid with $\beta=0.34$
(filled black diamonds). The mean zonal flows are estimated using
manual dye tracking for the full tanks and LDA measurements for the
1/2 tanks. The dashed line represents the
analytical solution in a sphere derived by \cite{sauretjfm2012}}{ZF_f05hz_various_dphi}

\clearpage
\pict[15cm]{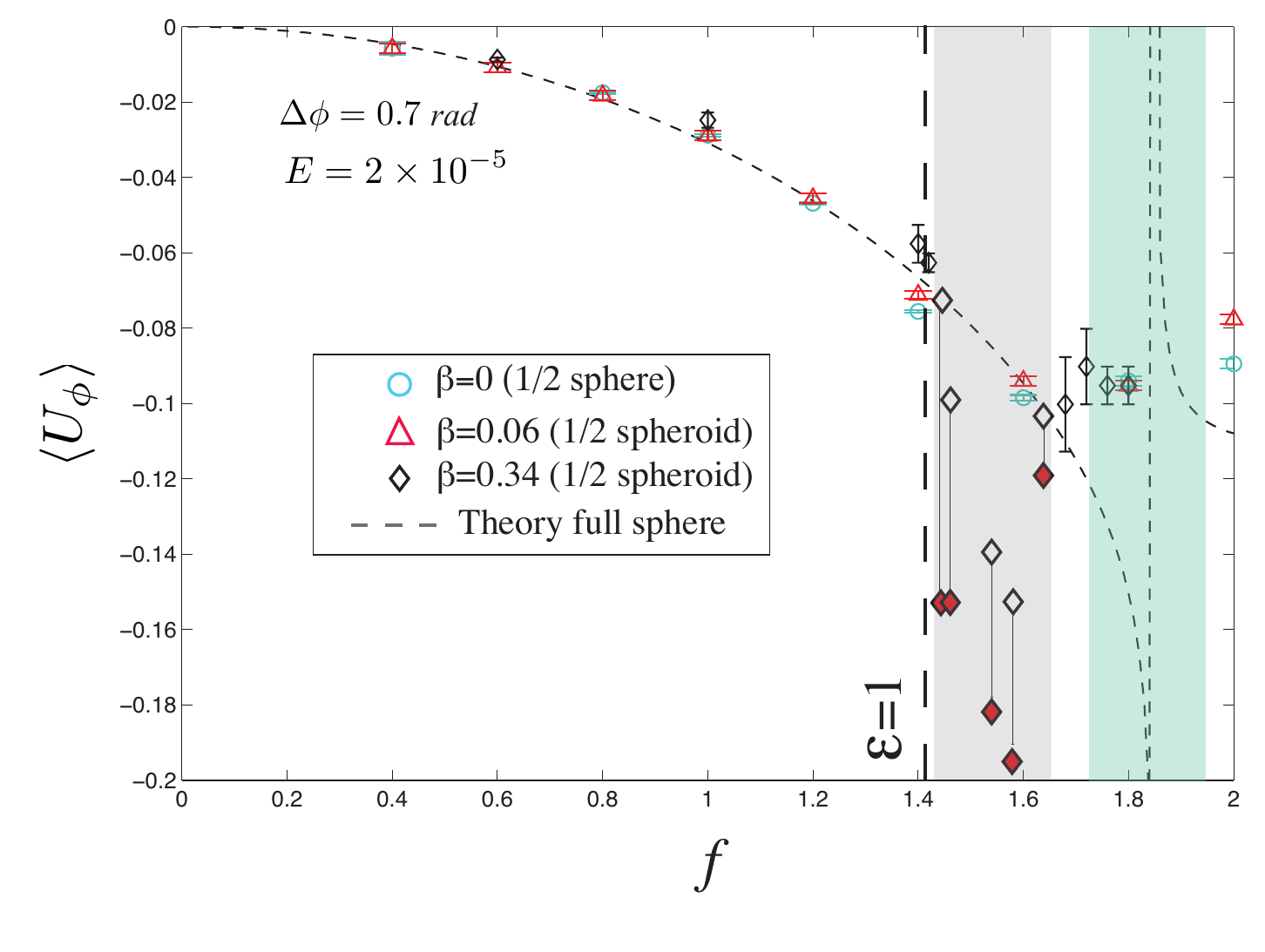}{Mean zonal flow from LDA
measurements as a function of the dimensionless libration frequency
for three different geometries; hemisphere (blue circle), 1/2
spheroid with $\beta=0.06$ (red triangle up) and 1/2 spheroid with
$\beta=0.34$ (black diamonds). The light gray rectangle represents the frequency band $f \in
[1.43; 1.66]$ for which we
observe laminar-turbulence intermittency at $\beta=0.34$. In such cases, we
distinguish between the zonal flow during the turbulent phases (red
filled diamond) and the zonal flow during the laminar phases (open
diamonds). The dashed line represents the
analytical solution in a sphere derived by \cite{sauretjfm2012}. The green rectangle represents the frequency range where the $E^{1/5}$ wide geostrophic shear structure influences the zonal flow measurements. In this region the analytical model represented by the dashed line is not expected to be valid.}{ZF_dphi80_various_f}

\clearpage
\pict[12cm]{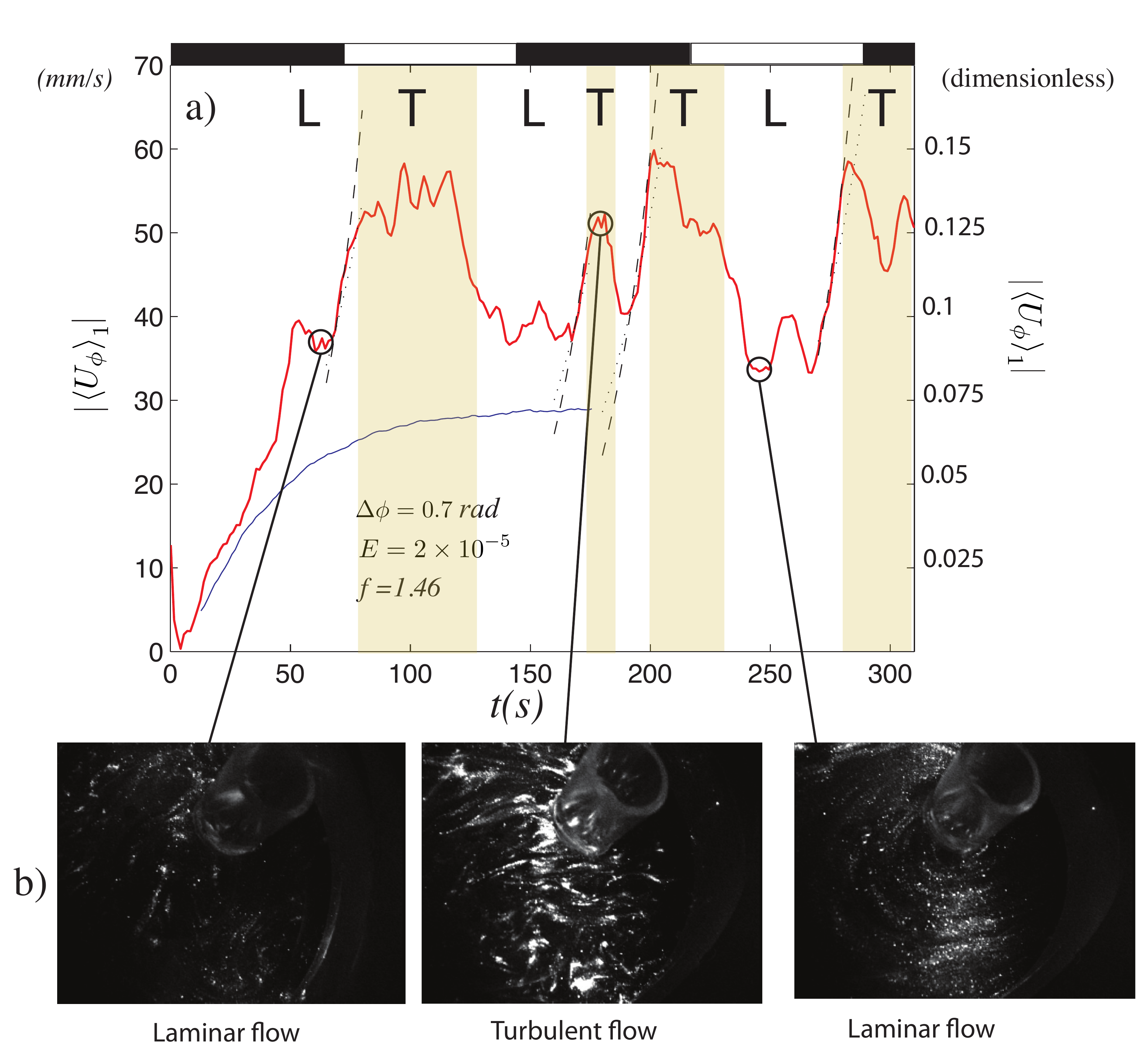}{ a) Time evolution of
the norm of the azimuthal velocity averaged over 10 oscillations for $\Delta\phi=0.7 rad$ ($\varepsilon
\sim1$) in the 1/2 spheroid with $f=1.46$, $\beta=0.34$ (red) and $\beta=0.06$ (blue). The measurements are performed at a cylindrical radius $S_i=48mm$ along the short axis of the mean equatorial ellipse, 1cm below the top flat surface. We perform a sliding window averaging over 10 oscillations with an overlap of 90$\%$. In addition we represent the WKB exponential growth for two extreme values of the dissipation factor, $K=1$ (dotted black) and $K=10$ (dashed black) as predicted by \cite{cebron2012letter}. The letters L and T stand for Laminar and Turbulent. The periods of turbulence, as observed in direct visualizations, are qualitatively represented by the yellow bands. Each top black and white rectangle represents a spinup time ($\sim 70$s). 
b) Top views of the shear structures in a plane
parallel to the equator. The time stamp of each snapshot is
indicated by circle in a). The first and last pictures show little
structures, which is characteristic of a laminar flow. In contrast,
the second snapshot exhibits numerous small scale structures,
typical of turbulent flows.}{ZF_30RPM_073HZ_80DEG_LDA_VIDEO}

\clearpage

\clearpage
\begin{sidewaystable}[h]
\begin{center}{\footnotesize
\begin{tabular}{|c|c|c|c|c|c|c|c|c|c|}
\hline
Planets&Internal layer&$r_o$(km)&$r_i$(km)&$T_{spin}$ (day)& $f^*$&$\Delta\phi $ (rad)&$E$&$\varepsilon$\\ 
\hline
Callisto$^{(1,2)}$& SO &$\sim$2300& 2000-2300&16.68&1& $4.22\times 10^{-6}$ & $4\times10^{-14}$ & $4.22\times 10^{-6} $\\
Ganymede$^{(3,4)}$&LC &$\sim$800&0-500&7.15&1&$5.64\times 10^{-6}$& $4\times10^{-14}$&$5.64\times 10^{-6}$\\
Earth's moon$^{(5)}$ &LC&$\sim$350&0-150&27.3&1&$7\times 10^{-5}$&$10^{-12}$&$7\times 10^{-5}$\\            
Titan(Grav) $^{(6,7)}$&SO&$\sim$2500&2350-2450&15.95&1&$2.3\times 10^{-5}$&$3.5\times10^{-14}$&$2.3\times 10^{-5}$\\
Mercury $^{(8)}$&LC&$\sim$1800&0-1700&58.6&2/3&$2\times 10^{-4}$ $^{(8)}$&$7.5\times10^{-14}$&$1.33\times 10^{-4}$\\
Titan(Atm)$^{(6,7)}$&SO&$\sim$2500&2350-2450&15.95&$3\times 10^{-3}$&$4.36\times 10^{-2}$ $^{(6)}$&$3.5\times10^{-14}$&$1.3\times 10^{-4}$\\
Io$^{(9)}$&LC&$\sim$500& - &1.77&1&$1.3\times 10^{-4}$&$3\times10^{-14}$&$1.3\times 10^{-4}$\\
Europa$^{(10)}$&SO&$\sim$1450&1300-1400&3.55&1&$2\times 10^{-4}$ $^{(10)}$&$.23\times10^{-14}$&$2\times 10^{-4}$\\
\hline
\end{tabular}}
\end{center}
\caption{Physical and dimensionless parameters values for planets, listed from top to bottom in terms of increasing boundary layer
Reynolds number value (Courtesy of \cite{noir09}). Titan(Grav) and Titan(Atm) correspond to Titan's forced longitudinal librations driven, respectively, by gravitational coupling and by atmospheric circulation. The anagram SO and LC stand for subsurface ocean and liquid metal core. Unless specified, the amplitudes of libration are from \citet{comstock03}. We use viscosities $\nu=10^{-6}$ and $\nu=3\time 10^{-7}$ m$^2/$s, respectively, for subsurface oceans and molten iron rich core. $T_{spin}$ is the rotational period of the planet. (1) \citet{Kuskov05}, (2) \citet{spohn03}, (3) \citet{hauk06}, (4) \citet{sohl02}, (5) \citet{williams02}, (6) \citet{lorenz08}, (7) \citet{tobie05}, (8) \citet{margot07}, (9) \citet{anderson01}, (10) \citet{hoolst08}}
\label{table_planets}
\end{sidewaystable}

 \end{document}